\documentclass[aps,twocolumn,nofootinbib,floatfix,superscriptaddress]{revtex4}
\usepackage{amsfonts}
\usepackage{amssymb}
\usepackage{amsmath}
\usepackage{graphics}
\usepackage{graphicx}
\usepackage{soul}
\usepackage{natbib}
\usepackage{bm}
\usepackage{mathtools}
\usepackage{stmaryrd}
\usepackage{epstopdf}
\usepackage{color}
\usepackage{lipsum}
\usepackage{balance}
\usepackage{enumitem}
\usepackage[colorlinks=true,citecolor=black,linkcolor=black,urlcolor=black,filecolor=black]{hyperref}

\usepackage{acronym}
\newacro{${1D}$}{one-dimensional}
\newcommand{\oneD}{\ac{${1D}$}}
\newacro{DF}{distribution function}
\newacroplural{DF}[DFs]{distribution functions}
\newcommand{\DF}{\ac{DF}}
\newcommand{\DFs}{\acp{DF}}
\newacro{HMF}{Hamiltonian Mean Field}
\newcommand{\HMF}{\ac{HMF}}
\newacro{BL}{Balescu--Lenard}
\newcommand{\BL}{\ac{BL}}
\newacro{FT}{Fourier transform}
\newcommand{\FT}{\ac{FT}}
\newacro{RK4}{fourth-order Runge-Kutta}
\newcommand{\RK}{\ac{RK4}}
\newacro{QSS}{quasi-stationary state}
\newacroplural{QSS}[QSSs]{quasi-stationary states}
\newcommand{\QSS}{\ac{QSS}}

\newacro{GL}{Gauss-Legendre}
\newcommand{\GL}{\ac{GL}}
\newacro{DoS}{density of state}
\newcommand{\DoS}{\ac{DoS}}

\newcommand{\rd}{\mathrm{d}}

\newcommand{\rr}{\mathrm{r}}
\newcommand{\rl}{\mathrm{l}}
\newcommand{\sech}{\mathrm{sech}}
\newcommand{\Card}{\mathrm{Card}}
\newcommand{\Sign}{\mathrm{Sign}}

\newcommand{\Mtot}{M_{\mathrm{tot}}}
\newcommand{\Etot}{E_{\mathrm{tot}}}
\newcommand{\Htot}{H_{\mathrm{tot}}}

\newcommand{\psid}{\psi^{\mathrm{d}}}
\newcommand{\deltaD}{\delta_{\mathrm{D}}}
\newcommand{\kB}{k_{\mathrm{B}}}
\newcommand{\rapo}{r_{\mathrm{a}}}
\newcommand{\Uper}{U_{\mathrm{per}}}
\newcommand{\ellmax}{\ell_{\mathrm{max}}}

\newcommand{\Tmax}{T_{\mathrm{max}}}
\newcommand{\Tbal}{T_{\mathrm{bal}}}
\newcommand{\tdyn}{t_{\mathrm{dyn}}}

\newcommand{\half}{\tfrac{1}{2}}
\newcommand{\kmax}{k_{\mathrm{max}}}

\newcommand{\dEbin}{\delta E_{\mathrm{bin}}}

\newcommand{\bI}{\mathbf{I}}
\newcommand{\bM}{\mathbf{M}}

\newcommand{\mC}{\mathcal{C}}
\newcommand{\mO}{\mathcal{O}}
\newcommand{\Flux}{\mathcal{F}}

\newcommand{\xp}{x^{\prime}}
\newcommand{\vp}{v^{\prime}}

\newcommand{\kp}{k^{\prime}}
\newcommand{\Jp}{J^{\prime}}
\newcommand{\up}{u^{\prime}}
\newcommand{\fp}{f^{\prime}}
\newcommand{\gp}{g^{\prime}}

\newcommand{\p}{\partial}
\newcommand{\dpd}[2][]{\frac{\p #2}{\p #1}}
\newcommand{\dd}[2][]{\frac{\rd #2}{\rd #1}}
\newcommand{\dds}[2][]{\frac{\rd^2 #2}{\rd #1^2}}

\usepackage{soul} \usepackage{xcolor}
\definecolor{aquamarine}{rgb}{0.5, 1.0, 0.83}

\begin{document}

\title{Long-term relaxation of ${1D}$ self-gravitating systems}

\author{Mathieu Roule}
\affiliation{Institut d'Astrophysique de Paris, UMR 7095, 98 bis Boulevard Arago, F-75014 Paris, France}
\author{Jean-Baptiste Fouvry}
\affiliation{Institut d'Astrophysique de Paris, UMR 7095, 98 bis Boulevard Arago, F-75014 Paris, France}
\author{Christophe Pichon}
\affiliation{Institut d'Astrophysique de Paris, UMR 7095, 98 bis Boulevard Arago, F-75014 Paris, France}
\affiliation{IPhT, DRF-INP, UMR 3680, CEA, L’Orme des Merisiers, Bât 774, 91191 Gif-sur-Yvette, France}
\author{Pierre-Henri Chavanis}
\affiliation{Laboratoire de Physique Th\'eorique, Universit\'e de Toulouse, CNRS, UPS, France}

\begin{abstract}
We investigate the long-term relaxation
of \oneD\@ self-gravitating systems,
using both kinetic theory and $N$-body simulations.
We consider thermal and Plummer equilibria,
with and without collective effects.
All combinations are found to be in clear
agreement with respect to the Balescu--Lenard and Landau predictions for the diffusion coefficients.
Interestingly, collective effects reduce the diffusion by a factor ${\sim\!10}$.
The predicted flux for Plummer equilibrium matches the measured one,
which is a remarkable validation of kinetic theory.
We also report on a situation of quasi kinetic blocking for the same equilibrium.
\end{abstract}
\maketitle

\section{Introduction}
\label{sec:Introduction}
The master equation describing the long-term evolution
of isolated discrete self-gravitating systems
is the so-called inhomogeneous \BL\@ equation~\citep{Heyvaerts2010,Chavanis2012}.
Such a formalism is particularly valuable because it captures
analytically some of the key non-linear processes
involved in these systems' orbit reshuffling
when driven by Poisson shot noise.
Yet, this kinetic framework relies on specific
sets of asymptotic assumption,
e.g.\@, timescale separation and sharp resonance conditions,
which may not be strictly fulfilled in practice.
Quantitative validation is therefore of interest.
Such assessments have been attempted
both for razor thin discs and spherical isotropic 
clusters~\citep[see, e.g.\@,][]{Fouvry2015,FouvryHamilton2021}.
However, the large dimension of phase space in these ${2D}$ and ${3D}$ systems
made these comparisons challenging,
as they involved intricate linear response
and summation of numerous resonances
over complex manifolds.
These works offered some qualitative agreement
between the kinetic predictions and simulations.
Yet, the quantitative accuracy of the match remained limited,
in particular because predictions require
repeated costly integrals over phase space,
while preserving long-term numerical precision
in the simulations is challenging.

This is the motivation for the present work,
which aims at performing such a thorough comparison
for one-dimensional self-gravitating systems,
whose reduced phase space dimension allows
for finer precision.
The long-term fate of \oneD\@ self-gravitating systems
was recently analysed numerically by~\cite{Joyce2010}.
Interestingly, such a model corresponds to a proxy
for more realistic astrophysical systems,
such as the vertical diffusion of stars~\citep[see, e.g.\@,][]{Solway+2012,Bovy+2012},
or the onset of large scale structure formation
in the early universe~\citep[see, e.g.\@,][]{Zeldovich1970,Valageas2006}.
Building upon~\cite{Benetti+2017},
which considered the long-term evolution of the \HMF\@
in its inhomogeneous phase,
the present investigation is also interesting
in what it shares or not with self-gravitating systems
of higher dimension.

Here, we aim at achieving a better understanding
of the mechanisms governing
the long-term evolution of discrete self-gravitating systems,
while accounting for collective effects (\BL\@) or not (Landau).
The paper is organized as follows.
Sec.~\ref{sec:Model} presents the model and the explored quasi equilibria.
Sec.~\ref{sec:Theory} computes their long-term resonant relaxation.
Sec.~\ref{sec:Discussion} explains the role of collective effects
and profile shapes 
in the system's long-term evolution, while 
Sec.~\ref{sec:Conclusions} sums up the lessons learned from this study case. 
All technical details are given in Appendices.

\section{Models}
\label{sec:Model}

\subsection{\oneD\@ self-gravitating systems}
\label{subsec:1DModel}

We consider a population of $N$ particles
of individual mass ${ m \!=\! \Mtot / N }$,
with $\Mtot$ the system's total mass.
Particles are confined to an infinite line
and coupled to one another via the pairwise interaction potential
\begin{equation}
U (x , \xp) = G \, |x - \xp| ,
\label{def_U}
\end{equation}
with $G$ the gravitational constant, and ${(x,\xp)}$
the respective positions of the two interacting particles.
This interaction corresponds to 
infinite parallel planes of uniform surface mass density 
attracting one another through the classical ${3D}$ Newtonian interaction.
The potential, $\psi (x)$, and density, $\rho(x)$,
are linked via Poisson's equation
\begin{equation}
    \Delta \psi = 2 G \rho\, ,
    \label{derivative_Poisson_eq}
\end{equation}
making the force between two particles
independent of their separation.
The \oneD\@ gravitational potential
differs from its ${3D}$ counterpart
in two respects:
(i) it is unbounded at large separation,
hence all particles are trapped (i.e.\ no escapers are possible);
(ii) it is finite at zero separation
allowing particles to cross one another.

Following an initial violent relaxation~\citep{LyndenBell1967},
the system's mean state can be described
by its ensemble-averaged \DF\@, ${ F \!=\! F (x , v , t) }$,
with $v$ the velocity, and normalized so that ${ \!\int\! \rd x \rd v F \!=\! \Mtot }$.
As all \oneD\@ equilibria are integrable, 
such a \QSS\ can most efficiently be described
via the angle-action coordinates ${(\theta,J)}$,
with $J$ the action,
and $\theta$ the associated ${2\pi}$-periodic angle
(see Appendix \ref{subapp:AA} for details).
In the absence of perturbations,
the angle evolves linearly in time
with the orbital frequency ${\Omega (J) \!=\! \p H/ \p J}$
where ${H \!=\! v^2 / 2 \!+\! \psi(x)}$ is the specific energy
and ${\psi(x)\!=\!\!\int\! \rd\xp\rd\vp F(\xp,\vp) U(x,\xp)}$
the system's mean-field potential.
In the following, we use equivalently $J$
or the specific (unperturbed) energy $E$
to label orbits.

As a result of potential fluctuations induced by the finite number of particles,
this \QSS\@, ${ F \!=\! F(J,t) }$,
undergoes a slow and irreversible long-term relaxation,
captured by the inhomogeneous \BL\@ equation~\cite{Heyvaerts2010,Chavanis2012}.
Testing this prediction is the focus of this work.

\subsection{Thermodynamic and quasi stationary equilibria}
\label{subsec:GTE}

We consider two explicit distributions:
(i) the global thermodynamical equilibrium;
and (ii) a more peaked \QSS\@, analog of the ${3D}$ Plummer sphere,
as we now detail.

Unlike their ${3D}$ analogs,
\oneD\@ self-gravitating systems
have a well-defined
maximum entropy equilibrium state.
Under the constraints
of fixed total mass and energy,
its density follows~\citep{Spitzer1942,Camm1950,Rybicki1971,Joyce2010}
\begin{equation}
    \rho (x) = \frac{\Mtot}{2\Lambda}\; \sech^{2}\!\left( x / \Lambda \right) ,
    \label{eq:Thermal_density}
\end{equation}
with $\Lambda$ the system's characteristic length
(see Appendix \ref{subapp:QSS} for the associated potential),
while its \DF\ reads
\begin{equation}
    F(E) = \frac{2\Mtot}{\sqrt{\pi}\sigma\Lambda}\;\exp\!\left(-2 E / E_{0} \right) ,
    \label{eq:Thermal_DF}
\end{equation}
with ${ \sigma \!=\! \sqrt{G\Mtot\Lambda} }$,
and ${ E_{0} \!=\! G \Mtot \Lambda }$
the characteristic velocity and specific energy.
We emphasize that the \DF\ from Eq.~\eqref{eq:Thermal_DF}
cannot further relax by design.
Naturally, this does not prevent individual particles
from undergoing themselves a diffusion.

We also investigate an equilibrium
stemming from polytropes~\citep{Eddington1916,Henon1973,Horedt2004}.
More precisely, by analogy with the ${3D}$ Plummer sphere,
we consider the \oneD\@ density
\begin{equation}
    \rho (x) = \frac{\Mtot}{2\alpha}\left[1+\left(x / \alpha \right)^2\right]^{-3/2} ,
    \label{eq:Plummer_density}
\end{equation}
where ${ \alpha \!=\! 2 \Lambda / \pi }$
ensures that this distribution has the same energy as Eq.~\eqref{eq:Thermal_density}.
The associated \DF\@ follows the power law distribution
(see Appendix \ref{subapp:QSS})
\begin{equation}
    F(E) = \frac{15\,G^3\,\Mtot^4\,\alpha^2}{32\sqrt{2}} \, E^{-7/2} .
    \label{eq:Plummer_DF}
\end{equation}

\begin{figure}[htbp!]
\begin{center}
\includegraphics[width=0.43\textwidth]{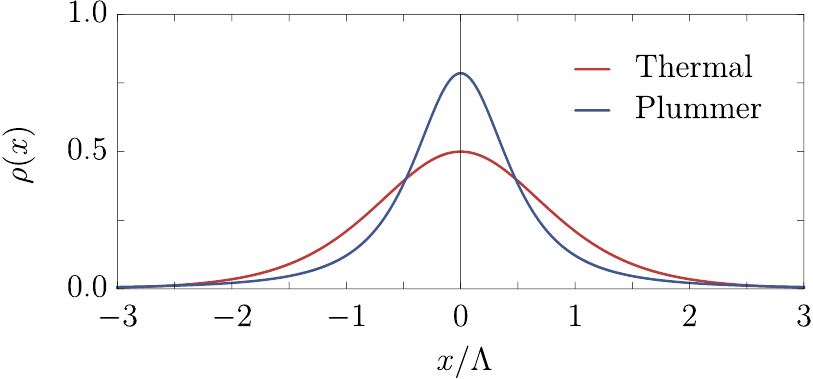}
\includegraphics[width=0.43\textwidth]{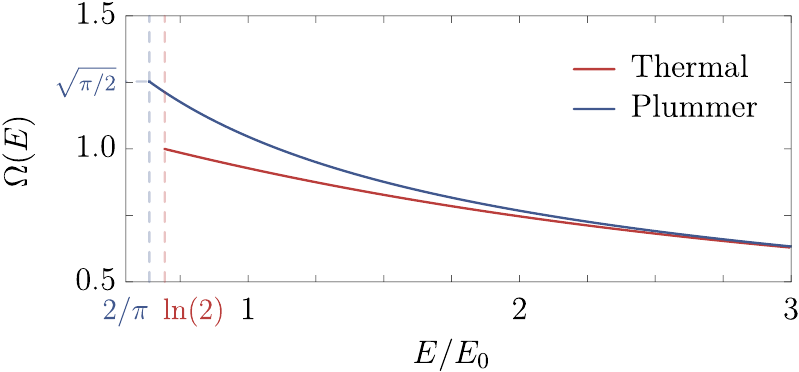}
\caption{{Top:} Density profiles of the thermal and Plummer equilibria. 
The Plummer equilibrium has a sharper core.
{Bottom:} Corresponding frequency profiles.
The range of available frequencies
is wider for the Plummer equilibrium.
\label{fig:densityprofiles}}
\end{center}
\end{figure}

In Fig.~\ref{fig:densityprofiles}, we illustrate
the density and frequency profiles
of these two states.
While the thermodynamical equilibrium has a strong core 
and few particles in the tails (only ${\!\sim\!10^{-9}}$ of the total mass outside 
${[-10\Lambda,10\Lambda]}$), 
the Plummer distribution
has a sharper core and much wider tails 
(${\sim\!10^{-3}}$ of the total mass outside 
${[-10\Lambda,10\Lambda]}$).
In the second panel of Fig.~\ref{fig:densityprofiles},
we present the frequency profile of both equilibria.
The Plummer denser core
widens its frequency profile,
allowing in turn for more resonances.
At high energies, both frequency profiles decrease like $1/\sqrt{E}$.

\section{Long-term Evolution}
\label{sec:Theory}

The long-term relaxation of self-gravitating systems
driven by finite-$N$ fluctuations
is generically governed by the inhomogeneous
\BL\@ equation~\citep{Heyvaerts2010,Chavanis2012}
\begin{align}
{} & \dpd[t]{F (J , t)} \!=\! - 2 \pi^{2} \, m \, \dpd[J]{} 
\!\bigg[ \sum_{k , \kp} k \!\! \int \!\! \rd \Jp  
\big| \psid_{k\kp} \big(J , \Jp , k \, \Omega (J)\big) \big|^{2}
\nonumber
\\
\times {} & \deltaD \big( k \, \Omega (J) \!-\! \kp \, \Omega (\Jp) \big) 
\bigg(\! \kp \dpd[\Jp]{} \!-\! k \dpd[J]{} \!\bigg) F (J) \, F(\Jp) \bigg] .
\label{BL_eq}
\end{align}
This non-linear equation describes the long-term evolution 
of the mean orbital distribution, ${F(J,t)}$,
driven by resonant couplings between gravitationally
dressed Poisson fluctuations (${ m \!\propto\! 1/N }$).
The sum, ${ \sum_{k,\kp} }$, and integral, ${ \!\int\! \rd \Jp }$,
in Eq.~\eqref{BL_eq} correspond
to a scan over the discrete resonances and 
orbital space.
Any time the resonance condition,
${k \, \Omega (J) \!-\! \kp \, \Omega (\Jp)\!=\!0}$,
is met, the diffusion is sourced.
The system's propensity to amplify fluctuations
is captured in the dressed susceptibility coefficients,
${|\psi^{\rd}_{k\kp}(J,\Jp,k\,\Omega)|^{2}}$.
Those are the (squared norm of the) \FT\@
of the pairwise interaction potential 
dressed by the system's gravitational susceptibility
(see Appendix \ref{subapp:FT}). 
In the following, we investigate both cases
where the gravitational dressing
is (\BL\@) and is not (Landau)
taken into account.

\subsection{Orbital Diffusion}
\label{subsec:Diffusion}

The \BL\@ Eq.~\eqref{BL_eq} can be re-written as a 
more compact continuity equation in action space
\begin{subequations}
\begin{align}
    \label{BL_eq_flux}
    \dpd[t]{F} &=   - \dpd[J]{\Flux}  \\
    \label{BL_eq_diff_fric}
    &= - \dpd[J]{} 
    \left[
        A(J) F(J) - \half \, D(J) \dpd[J]{F}
    \right],
\end{align}
\label{BL_eq_def_flux}
\end{subequations}
with the total flux ${ \Flux (J , t) }$, and the diffusion coefficient
\begin{align}
    \nonumber D(J) =  (2\pi)^2 m {} & 
    \sum_{k,\kp} k^2 \!\! \int \!\! \rd\Jp \,
    \big| \psid_{k\kp} \big(J , \Jp , k \, \Omega (J)\big) \big|^{2} \\
    {} & \times \deltaD \big( k \, \Omega (J) \!-\! \kp \, \Omega (\Jp) \big) \, 
    F(\Jp).
    \label{BL_diffusion_coefficient}
\end{align}
In Eq.~\eqref{BL_eq_diff_fric}, 
the polarization friction, ${A(J)}$, is obtained from Eq.~\eqref{BL_diffusion_coefficient}
via the substitutions
 ${(2\pi)^2\!\to\!2\pi^2}$, ${k^2\!\to\!k\,\kp}$ and 
${F\!\to\!\p F / \p \Jp}$.
As discussed in Sec.\@ 7.4.2 of~\cite{BinneyTremaine2008},
the diffusion coefficient also has the simple interpretation
\begin{equation}
D (J) = \underset{T \to + \infty}{\lim}
\frac{\left\langle  \Delta J^{2} (T)  \right\rangle}{T} ,
\label{diffusion_coefficient_interpretation}
\end{equation}
with ${ \Delta J (T) \!=\! J (t \!=\! T) \!-\! J (t \!=\! 0) }$
the change in action of a given particle,
and ${ \langle \cdot \rangle }$ the ensemble average
over realisations.
Equations~\eqref{BL_diffusion_coefficient} and \eqref{diffusion_coefficient_interpretation}
provide us with two independent means of measuring and predicting ${ D(J) }$.
In the following, we will focus our interest on the diffusion coefficients
in energy, which naturally read ${D_{EE} \!=\! \Omega^2 D}$.

\subsection{Diffusion coefficients}
\label{subsec:coef}

\begin{figure}[htbp!]
\begin{center}
\includegraphics[width=0.48\textwidth]{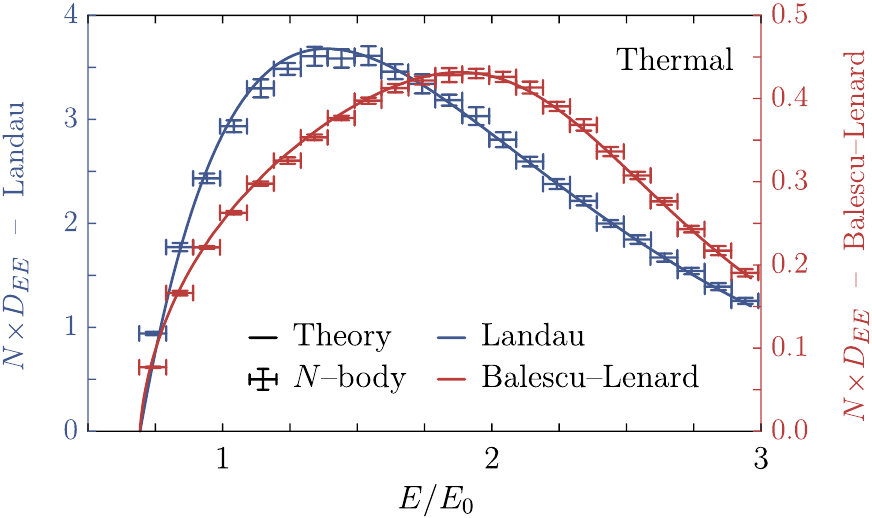}
\includegraphics[width=0.48\textwidth]{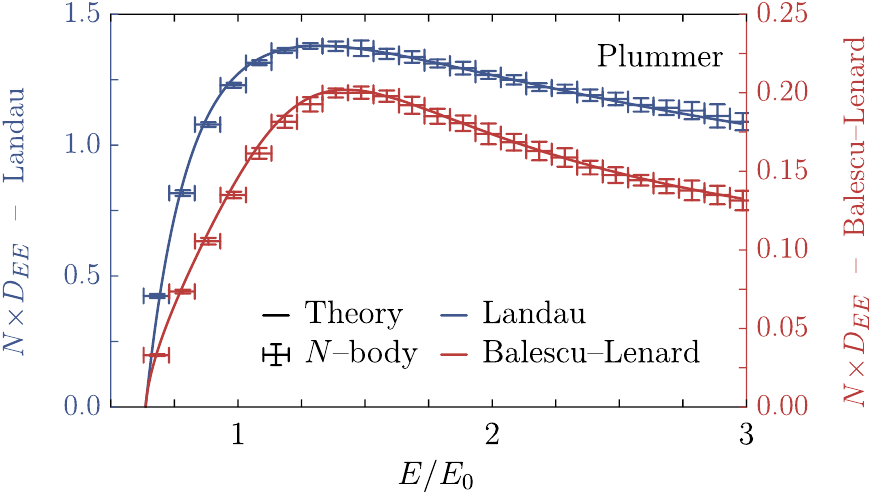}
\caption{Top: Diffusion coefficients at thermal equilibrium as a function of energy 
in both Landau (i.e.\ without collective effects) and
\BL\@ (i.e.\ with collective effects) cases.
Bottom: Same as the top panel but for the Plummer equilibrium.
The kinetic theory shows a very satisfactory 
match to the numerical measurements.
For illustration, both measurements have their 
own adapted scale as 
collective effects slow down diffusion by a factor ${\sim\!10}$.
See Appendix \ref{subapp:diff_measure} for the numerical details.
\label{fig:DEE}}
\end{center}
\end{figure}

In the top panel of Fig.~\ref{fig:DEE},
we present the diffusion coefficients at thermal equilibrium computed with the 
\BL\@ and the Landau formalism, together
with the corresponding estimates from numerical simulations.
We refer to Appendix \ref{app:technical} for the details
of the kinetic estimation,
and Appendix \ref{app:NBodyIntegration} for the $N$-body measurements.
In both Landau and \BL\@ cases, we recover a very good match
between the kinetic theory and the numerical measurements. 
This confirms that, indeed, long-range resonant couplings
are responsible for the long-term relaxation of these systems.
We stress that the \BL\@ diffusion coefficients 
are ${\sim\!10}$ times smaller than the 
Landau ones,
an effect already noted in the \HMF\@ model 
for highly magnetized thermal equilibria~\citep[see fig.~{9} in][]{Benetti+2017}.
This is at variance with the low magnetization HMF result,
or the case of self-gravitating stellar disks 
\cite{Fouvry2015} where collective effects considerably accelerate the relaxation.

In the bottom panel of Fig.~\ref{fig:DEE},
we present the same measurements for the Plummer distribution.
Satisfactorily, this other equilibrium shows
the same level of fine agreement.
Similarly, we also find that
collective effects slow down the diffusion by a factor ${\sim\!10}$.
This will be discussed in Sec.~\ref{sec:Discussion}.

\subsection{Fluxes}
\label{subsec:Flux}

\begin{figure}[htbp!]
    \begin{center}
    \includegraphics[width=0.43\textwidth]{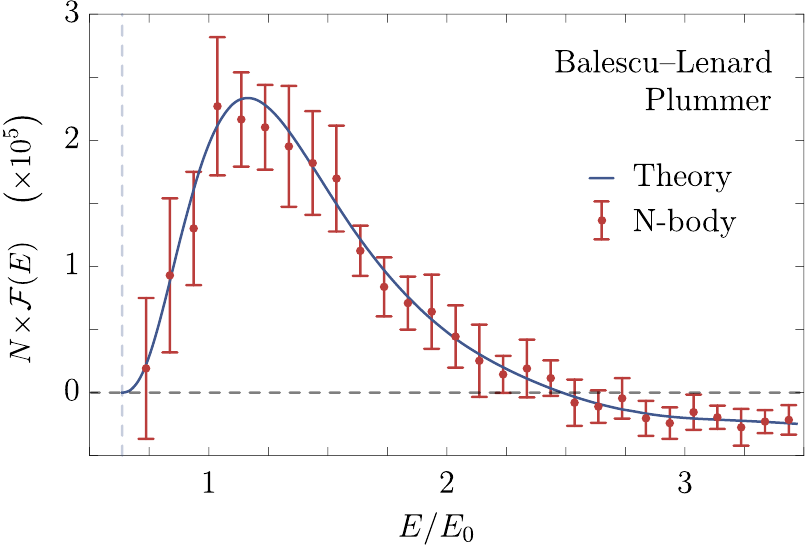}
    \caption{Initial flux in the Plummer equilibrium as a function of energy predicted by  kinetic theory (line) and 
    measured in $N$-body simulations (points).
    Both capture the maximum near ${ E / E_{0} \!\simeq\! 1.25 }$,
    and the change of sign at ${ E/E_{0} \!\simeq\! 2.5 }$.
    We stress that the amplitude of ${ N \!\times\! \Flux(E) }$
    has been rescaled by $10^{5}$.
    See Appendix \ref{subapp:flux_measure} for the numerical details.
    \label{fig:Flux}}
    \end{center}
\end{figure}

We now turn our interest to the initial diffusion flux,
${ \Flux (J , t\!=\!0) }$,
as given by Eq.~\eqref{BL_eq_def_flux}.
Of course, this flux vanishes for the thermodynamical equilibrium.
In Fig.~\ref{fig:Flux}, we illustrate the initial diffusion flux
for a fully self-gravitating Plummer equilibrium.
Once again, the kinetic theory and numerical simulations
are found to be in a good match,
and both recover the (slow) relaxation of the Plummer distribution
towards the thermal one.
Within the appropriate dimensionless units,
we point out that the diffusion flux is typically ${\sim\!10^{5}}$ times smaller
than the diffusion coefficients,
i.e.\ the efficiency of the relaxation is drastically hampered
by a near kinetic blocking.
This is further discussed in Sec.~\ref{subsec:Freq_profile}.

\subsection{Correlation of the perturbations}
\label{sec:Timecorr}

\begin{figure}[htbp!]
    \begin{center}
    \includegraphics[width=0.43\textwidth]{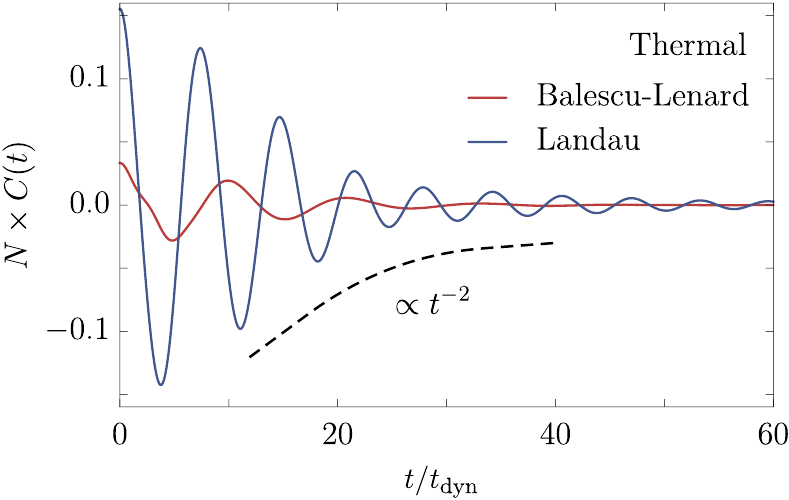}
    \caption{Time correlation, ${ C (t) \!=\! \langle \delta \psi (0) \, \delta \psi (t) \rangle }$
    of the potential fluctuations in $N$-body simulations
    of the thermal equilibrium
    with and without collective effects,
    as a function of the time ${ t / \tdyn }$.
    See Appendix \ref{subapp:TC_measure} for precise definitions.
    In the presence of collective effects,
    both the amplitude and coherence time of the correlation fonction
    are reduced.
    \label{fig:TC_thermal}}
    \end{center}
    \end{figure}

Following~\cite{Fouvry2018},
we present in Fig.~\ref{fig:TC_thermal}
the correlation ${ C (t) \!=\! \langle \delta \psi (0) \, \delta \psi (t) \rangle }$ of the potential fluctuations, ${ \delta \psi (t) }$,
in the $N$-body simulations,
as a function ${ t / \tdyn }$,
with ${ \tdyn \!=\! \Lambda / \sigma }$, the dynamical time. This correlation sources orbital diffusion~\citep{Binney+1988}.
We refer to Appendix \ref{subapp:TC_measure} for a precise
definition of ${ C (t) }$.
The gravitational dressing has two main effects:
(i) it weakens the overall amplitude of the potential fluctuations;
(ii) it reduces the coherence time of these perturbations.
Naturally, this drives a slower orbital diffusion
in the \BL\@ situation compared to the Landau one,
as presented in Sec.~\ref{subsec:Diffusion}.

\begin{figure}[htbp!]
    \begin{center}
    \includegraphics[width=0.48\textwidth]{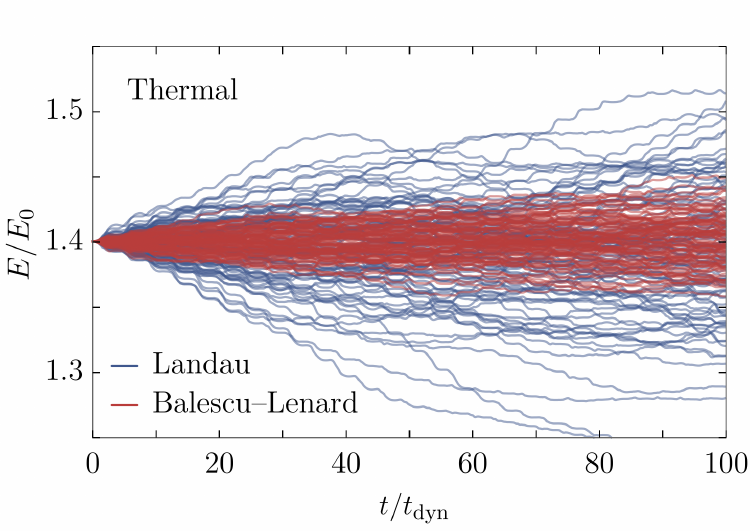}
    \caption{Typical diffusion of test particles
    embedded within $N$-body realizations of the thermal equilibrium
    with collective effects (\BL\@) or without (Landau).
    The massless test particles are all placed at the same initial
    phase space location
    in their respective realizations.
    Collective effects slow down the orbital diffusion.
    \label{fig:diff_BL_Landau}}
    \end{center}
    \end{figure}

This is fully consistent with Fig.~\ref{fig:diff_BL_Landau}
where we equivalently illustrate the diffusion 
of individual test particles
in the presence/absence of collective effects.
In that figure, we also recover that the energy diffusion
is naturally modulated at the frequency ${\sim2\pi/\tdyn}$,
i.e.\ the typical frequency of the background thermal equilibrium.

\section{Discussion}
\label{sec:Discussion}

We now discuss our two main findings:
non-thermal equilibria present very inefficient relaxation;
and collective effects reduce the efficiency of diffusion.

\subsection{Quasi kinetic blocking}
\label{subsec:Freq_profile}

In Fig.~\ref{fig:Flux}, we noted that, within appropriate dimensionless units,
the diffusion flux in the Plummer equilibrium
is ${\sim\!10^{5}}$ times smaller than the associated
diffusion coefficients (see Fig.~\ref{fig:DEE}).
This is the imprint of a (quasi-) kinetic blocking,
highlighting the system's difficulty to populate resonances
driving an efficient diffusion.

As put forward in Eq.~\eqref{BL_eq},
the system's long-term diffusion is sourced by resonant interactions.
For a given resonant pair ${ (k,\kp) }$,
one has to ensure that the resonance condition,
${ k \Omega(J) \!-\! \kp \Omega(\Jp) \!=\! 0 }$,
is met, while the overall efficiency of this coupling
is governed by the susceptibility coefficients,
${ \psi_{k\kp}^{\rd} (J , \Jp , \omega) }$, for that pair.
In practice, a couple of important ``conspiracies'',
responsible for the small flux observed in Fig.~\ref{fig:Flux}, operate:
\begin{enumerate}[wide, labelindent=0pt, label=(\roman*)]
\setlength\itemsep{-0.1em}
\item The Plummer frequency profile is monotonic
(see Fig.~\ref{fig:densityprofiles}).
Any resonance ${ k \!=\! \kp }$
systematically imposes ${ J \!=\! \Jp }$,
leading to an exactly vanishing flux in Eq.~\eqref{BL_eq}.
\item Symmetry imposes ${ \psi_{k\kp}^{\rd} \!=\! 0 }$,
for all ${k,\kp}$ of different parity (see Appendix \ref{subapp:Multipole}).
As a consequence, one must have ${ |k \!-\! \kp| \!\geq\! 2 }$
for a resonance to contribute to the flux.
Similarly, ${k,\kp}$ must also have the same sign.
\item Despite its denser core, the overall frequency range
of the Plummer profile is still finite
(see Fig.~\ref{fig:densityprofiles}).
For a given orbit $J$, this imposes
${ k/\kp \leq \Omega(J\!=\!0) / \Omega (J) }$
for the resonance condition from Eq.~\eqref{BL_eq} to be met.
\item For $k$ large enough,
the bare susceptibility coefficients asymptotically scale like
${ \psi_{kk}(J,J) \!\propto\! 1/k^{2} }$ (see Appendix \ref{subapp:Multipole}).
The higher order the resonance,
the less efficient the coupling,
and hence the (drastically) smaller the contribution to the flux.
\end{enumerate}

\begin{figure}[htbp!]
    \begin{center}
    \includegraphics[width=0.48\textwidth]{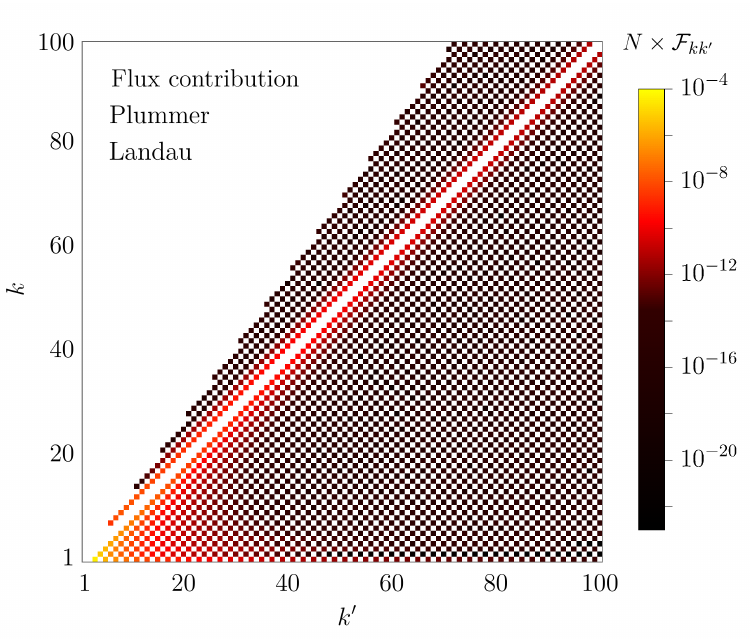}
    \caption{Individual contributions of the various resonances ${ (k,\kp) }$
    to the initial Landau flux, ${ \Flux (E , t \!=\! 0) }$,
    for the Plummer equilibrium
    and ${ E \!=\! \psi (2 \alpha) }$. By symmetry,
    we only consider ${k,\kp\!\geq\!0}$ resonances.
    Note the logarithmic colour coding.
    The flux is dominated by low order resonances and suffers from many 
    annihilating conspiracies (see main text).
    \label{fig:ContribRes}}
    \end{center}
\end{figure}

We highlight these different effects in Fig.~\ref{fig:ContribRes},
where we isolate the contributions, $\Flux_{k\kp} $, 
of the different resonances ${ (k,\kp) }$
to the Landau flux ${ \Flux \!=\! \sum_{k,\kp > 0} \Flux_{k\kp} }$.
We emphasize in particular the rapid decay of the flux contributions
as ${ k,\kp }$ increase
and as one moves away from the diagonal ${ k\!=\! \kp }$
(which only contributes to the diffusion coefficient
and not the flux).
These different effects
are jointly responsible for the small flux reported
in Fig.~\ref{fig:Flux}.

Figure~\ref{fig:ContribRes} is essentially left
unchanged when taking into account collective effects.
The only significant difference in the \BL\@ case
is the reduced contribution from the resonances with ${k\!=\!1}$
for which gravitational dressing weakens the amplitude
of the orbital coupling as detailed 
in Sec.~\ref{subsec:LinearResponse} and ~\ref{subsec:DampingDressing}. 
Taking collective effects into account 
therefore further reduces the flux as they notably damp 
contribution from the resonance ${ (k,\kp) \!=\! (1,3)}$,
the main contributor to the Landau flux (see Fig.~\ref{fig:ContribRes}).

Despite this relative inefficiency,
we stress that the Plummer equilibrium
still relaxes through ${1/N}$ two-body resonant effects.
This is in stark contrast with homogeneous \oneD\@ systems
which are generically kinetically blocked at order ${1/N}$~\cite[see, e.g.\@,][]{Chavanis2012}
and require the derivation of appropriate kinetic equations
at order ${1/N^{2}}$ sourced by three-body effects~\citep{Fouvry2020Nsquared}.

\subsection{Linear Response}
\label{subsec:LinearResponse}

We now discuss the influence of collective effects.
The efficiency of the gravitational dressing
of perturbations is generically captured
by the response matrix, ${ \bM (\omega) }$~\citep[see, e.g.\@, Eq.~{(5.94)} in][]{BinneyTremaine2008}
which reads here
\begin{equation}
\bM_{pq} (\omega) = 2 \pi \sum_{k \in \mathbb{Z}} \! \int \! \rd J \, \frac{k \, \p F / \p J}{\omega \!-\! k \, \Omega (J)} \, \psi^{(p)*}_{k} \!(J) \, \psi^{(q)}_{k} (J) ,
\label{def_M}
\end{equation}
with ${ \psi^{(p)}_{k} (J) }$ the \FT\@ of
the bi-orthogonal basis elements.
As detailed in Appendix \ref{subapp:BOB},
we construct
natural basis elements by periodizing the interaction potential
on a \textit{ad hoc} length $L$.
Such a modification impacts the system
only on large separations (i.e.\ small frequencies),
which we alleviate by picking $L$ sufficiently large
given the system's density.
We refer to Appendix \ref{subapp:RepMat}
for details on the computation of the response matrix,
in particular regarding the resonant denominator
from Eq.~\eqref{def_M}.

\begin{figure}[htbp!]
\begin{center}
\includegraphics[width=0.43\textwidth]{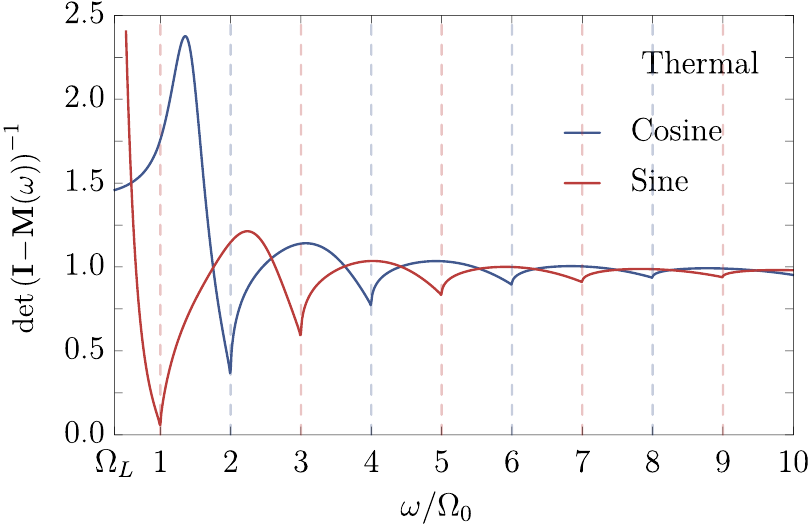}
\caption{
Determinant of the susceptibility matrix,
${ [\bI \!-\! \bM (\omega)]^{-1} }$,
as a function of the
real frequency ${ \omega / \Omega_{0} }$
for the even (cosine) and odd (sine) basis elements
(see Appendix \ref{subapp:BOB}).
Here, $\Omega_{0}$ is the maximum frequency
in the system's center,
while ${\Omega_L\!\simeq\!0.35\,\Omega_0}$ is the smallest frequency captured 
by the $L$-periodized potential (${L\!=\!10\,\Lambda}$). 
Collective effects become negligible at small separation (high frequencies).
Conversely, they induce a striking damping for frequencies
${\omega\!\sim\!\Omega_{0}}$,
which explains the particular inefficiency of the \BL\@ diffusion
compared to the Landau one.
\label{fig:Amplification}}
\end{center}
\end{figure}

In Fig.~\ref{fig:Amplification},
we illustrate the determinant
of the susceptibility matrix ${ [\bI - \bM (\omega)]^{-1} }$
for the thermal equilibrium,
as a function of ${ \omega / \Omega_{0} }$,
with ${\Omega_{0} \!=\! \sqrt{G\Mtot/\Lambda}}$
the (maximum) orbital frequency in the system's center
(${\Omega_{0} \!=\! \sqrt{G\Mtot/\alpha}}$ 
for the Plummer equilibrium).

Because the system possesses a finite maximum frequency,
$\Omega_{0}$, its linear response shows clear signatures
at every (resonant) multiple of this frequency.
Nonetheless, we find that the collective amplification
remains limited, while the same result also holds
for the Plummer equilibrium.
Conversely, collective effects significantly damp
the contribution of the odd resonances
${k\Omega\!\sim\!\Omega_{0}}$,
i.e.\ the lowest order resonances in the most populated regions.
These resonances being 
dominant contributors to the diffusion (see Sec.~\ref{subsec:DampingDressing}), 
it explains the relative inefficiency of the \BL\@ diffusion
unveiled in Fig.~\ref{fig:DEE}.
This is in sharp constrast with
${ \ell \!=\! 1 }$ perturbations in globular clusters~\citep[see, e.g.\@, fig.~{1} in][]{FouvryPrunet2022}.

\subsection{Impact of collective effects}
\label{subsec:DampingDressing}

The influence of the gravitational dressing
strongly depends on the resonance frequency,
${ \omega \!=\! k \Omega }$.
It is therefore of interest 
to pinpoint the individual contributions of resonances
to the diffusion coefficient, ${ D \!=\! \sum_{k,\kp>0} D_{k\kp} }$.

\begin{figure}[htbp!]
    \begin{center}
    \includegraphics[width=0.45\textwidth]{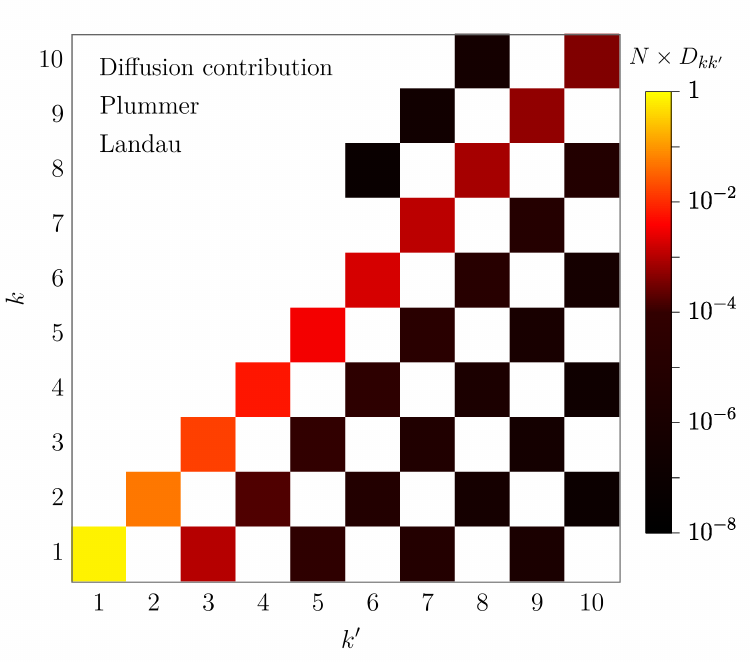}
    \includegraphics[width=0.45\textwidth]{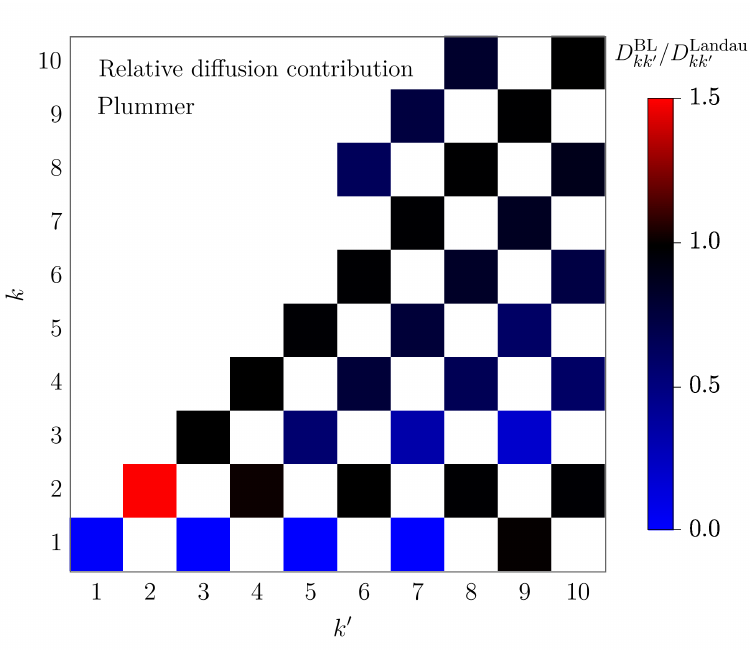}
    \caption{Top: Individual contributions of the various resonances ${ (k,\kp) }$
    to the Landau diffusion coefficients for the Plummer equilibrium
    and ${ E \!=\! \psi (2 \alpha) }$.
    Bottom: Relative contributions
    when collective are or not taken into account,
    for the same setup.
    The main contributor to the Landau diffusion, resonance ${(1,1)}$, 
    is severely damped by collective effects,
    while the amplification of other resonances remains limited. 
    \label{fig:Rescontrib_BL_vs_Landau}}
    \end{center}
\end{figure}

As for the flux, the allowed resonances must satisfy 
a parity criterium as well as ${ k/\kp \leq \Omega(J\!=\!0) / \Omega (J) }$,
but, however, ${k\!=\!\kp}$ resonances contribute to the diffusion.
Given that the coupling efficiency rapidly drops
with the order of the resonance,
in Fig.~\ref{fig:Rescontrib_BL_vs_Landau},
we focus on the contributions of low-order resonances.
The top panel of this figure
illustrates the predominant role of the resonance ${ (k,\kp) \!=\! (1,1)}$
in the Landau orbital diffusion (in yellow), while 
the bottom panel shows the extinguishing role of collective effects 
for any ${k\!=\!1}$ resonances.
This is ultimately responsible
for the relative inefficiency of the \BL\@ diffusion
w.r.t.\ the Landau one.
The determinant of the susceptibility matrix plotted in 
Fig.~\ref{fig:Amplification} allows us to
reach the same conclusions.
Indeed, the gravitational susceptibility
suffers from a drought for any odd resonant couplings with ${ \omega\!\sim\!\Omega_{0} }$.
And, the slight amplification
of the resonance ${(2,2)}$ observed in Fig.~\ref{fig:Rescontrib_BL_vs_Landau}
is equivalently found in Fig.~\ref{fig:Amplification}
since ${2\Omega(2\alpha)\!\sim\!1.4\Omega_0}$.
This amplification remains still too limited to compensate
for the strong collective damping of the dominating ${ (1,1) }$ resonance.

\section{Conclusions}
\label{sec:Conclusions}

The long-term relaxation of discrete self-gravitating systems
is driven by the subtle combined effects
of finite-${N}$ Poisson fluctuations and
long-range orbital resonances,
possibly boosted or damped by gravitational polarization.
This is captured by the inhomogeneous
\BL\@ equation~\citep{Heyvaerts2010,Chavanis2012}.
In this work, we compared its kinetic predictions
with $N$-body simulations of \oneD\@ self-gravitating systems.

We focused on the thermal and Plummer equilibria,
while accounting and not accounting for collective effects.
We reach clear agreement for both models
on the rate of diffusion. 
The \BL\@ diffusion coefficients were found
to be ${\sim\!10}$ times smaller than the Landau ones,
i.e.\ collective effects
surprisingly mitigates diffusion, and we provided an explanation 
for it.
This conclusion is particularly interesting
as it is also present in the \HMF\@ model
in highly magnetized equilibria~\citep{Benetti+2017}
while it is absent in weakly magnetized ones
or in the periodic stellar cube~\citep{Magorrian2021}. 
This may or may not be the case in higher dimensions~\citep[see, e.g.\@,][]{Weinberg1989,Fouvry2015,FouvryHamilton2021}
possibly depending on the position and geometry of the wake and on these systems' reservoirs
of free energy via rotation or anisotropy.

Similarly, the predicted flux closely matches the measured one
for the Plummer equilibrium.
This is a remarkable validation of kinetic theory,
which was not granted \textit{a priori} since the \BL\@ theory
makes strong assumptions about the amplitude of the fluctuations 
and timescale decoupling between the linear and long-term processes.
We discussed how diffusion is mostly driven
by low order resonances which can be significantly altered 
by collective effects. We explained how the vanishing contribution 
of ${k\!=\!\kp}$ resonances to the flux leads to a quasi-kinetic blocking,
drastically slowing down the relaxation 
of non-thermal equilibria.

Beyond this work, one should aim
at better understanding the precise origin of the ability
of collective effects to accelerate/slow down relaxation.
For example, one could investigate sets
of equilibria closer to marginal stability,
e.g.\@, with bumps on tail,
and identify the possible importance of their damped modes~\citep[see, e.g.\@,][]{Weinberg1994}.
In the spirit of the low magnetization \HMF\@ model~\citep{Benetti+2017},
one may expect that collective modes
would ultimately boost the \BL\@ flux over the Landau one.

Given the accuracy achieved for the initial flux, it
would clearly be useful to integrate
Eq.~\eqref{BL_eq} self-consistently in time.
This is no easy undertaking,
as it involves tracking both
the non-linear dependence in $F$ and the 
joint evolution of the mean potential
and the associated angle-action coordinates~\citep[see, e.g.\@,][]{Weinberg2001}.
This same \oneD\@ model may also prove useful
to understand the relaxation of thickened galaxies~\cite[see, e.g.\@,][]{Fouvry2017}.
More generally, it bodes well for future implementations in higher dimensions, as in globular clusters, or dark matter halos.

\begin{acknowledgments}
This work is partially supported by the grant Segal ANR-19-CE31-0017 
of the French Agence Nationale de la Recherche  (\href{https://secular-evolution.org}{https://secular-evolution.org}),
and by the Idex Sorbonne Universit\'e  (\href{https://ipi-sorbonne-universite.fr}{https://ipi-sorbonne-universite.fr}).
We thank S.\ Rouberol for the smooth running
of the Infinity cluster,
where the simulations were performed. We thank K.\ Tep and M.\ Petersen for many stimulating discussions.
The codes underlying the present work are distributed online 
at: \href{https://github.com/MathieuRoule/odiBLe}{https://github.com/MathieuRoule/odiBLe}.
\end{acknowledgments}

\vskip 1cm

\appendix

\section{\oneD\@ kinetic theory}
\label{app:technical}
\subsection{Angle-action coordinates}
\label{subapp:AA}
Following Eq.~{(3.195)} of~\cite{BinneyTremaine2008},
the action of an orbit is the circulation of ${v}$
for one full radial oscillation. For 
an even mean-field potential ${ \psi (r \!=\! |x|) }$,
it simply reads
\begin{equation}
\label{def_J}
J = \frac{1}{\pi} \!\! \int_{-\rapo}^{\rapo} \!\! \rd x \, v = \frac{2 \, \sqrt{2}}{\pi} \!\! \int_{0}^{\rapo} \!\! \rd r \, \sqrt{\psi (\rapo) - \psi (r)},
\end{equation}
with $\rapo$ the orbit's apocenter, i.e.\ the maximum radius
reached during the particle's libration which satisfies ${E\!=\!\psi(\rapo)}$.
In the following sections, we equivalently use $\rapo$, $E$ and $J$
to label orbits.
The orbital frequency, ${ \Omega \!=\! \p H / \p J }$,
and the associated angle $\theta$, satisfying $\dot\theta=\Omega$, read
\begin{subequations}
    \label{def_Omega_theta}
    \begin{align}
        \label{def_Omega}
        \frac{1}{\Omega} & = 
        \frac{\sqrt{2}}{\pi} \int_{0}^{\rapo} \!\! \frac{\rd x}{\sqrt{\psi (\rapo) - \psi(x)}},
        \\
        \label{def_theta}
        \theta(x,\rapo) & = 
        \frac{\Omega(\rapo)}{\sqrt{2}}  \int_{\mC}  \frac{\rd \xp}{\sqrt{\psi (\rapo) - \psi (\xp)}} ,
    \end{align}
\end{subequations}
with ${\mC}$ the contour going from ${ x \!=\! - \rapo }$
up to the current position ${ x \!=\! x (\theta) }$
along the radial oscillation.
Therefore, the angle mapping is such that 
\begin{equation}
    x (\theta \!=\! 0) = - \rapo ;
    \;
    x (\tfrac{\pi}{2}) = 0 ;
    \;
    x (\pi) = \rapo ;
    \;
    x (\tfrac{3 \pi}{2}) = 0 .
    \label{map_angle_part}
\end{equation}
To cure the divergence of 
the integrand of Eq.~\eqref{def_Omega_theta}
for ${x\!\rightarrow\! \pm \rapo}$,
we perform the change of variables
${x \!=\! \rapo f(u)}$ towards an effective anomaly ${ -1 \!\leq\! u \!\leq\! 1 }$ 
satisfying ${ f (\pm) \!=\! 1 }$ and ${\fp(\pm 1)\!=\!0}$.
This change of variable must be: (i) explicit (no inversion needed), 
(ii) stable (to sample numerous nearby points), 
(iii) generic (must work for any analytic potential).
In practice, we use the polynomial anomaly 
${f(u)\!=\!u(\tfrac{3}{2}\!-\! \half u^2)}$~\cite{Henon1971}.

To address the 
arising ${0/0}$ limit in the integrand
\begin{equation}
    I(u,\rapo) = \frac{\rapo \fp(u)}{\sqrt{\psi(\rapo) - \psi(\rapo f(u))}} ,
    \label{integrand_anomaly}
\end{equation}
we use a second-order Taylor expansion in ${u\!\to\!\pm 1}$ and 
${\rapo / \Lambda \!\to\!0^{+}}$,
as soon as
${|1\!\pm\!u|\!\leq\!10^{-3}}$ or ${\rapo/\Lambda\!\leq\!10^{-3}}$.
Benefiting from this numerically stable approach,
the integrals
from Eqs.~\eqref{def_J} and~\eqref{def_Omega_theta}
are computed using Simpson's ${1/3}$-rule,
with $100$ uniforms intervals in ${ u \in [0,1] }$.

\subsection{Biorthogonal basis}
\label{subapp:BOB}
 
Following~\cite{Kalnajs1976II}, the bi-orthogonal basis
elements satisfy
\begin{subequations}
    \label{bi_ortho_basis}
    \begin{align}
        \label{Poisson_eq_rho_psi}
         & \psi^{(p)}(x) = \!\! \int \!\! \rd \xp \, \rho^{(p)}(\xp)\, U(x,\xp), {} & \\
        \label{bi_ortho_relation}
         & \!\! \int \!\! \rd x \, \rho^{(p)}(x) \, \psi^{(q)*}(x)  = - \delta_{pq}. {} & 
    \end{align}
\end{subequations}
With them, the pairwise interaction potential
becomes
\begin{equation}
    U(x,\xp) = - \sum_{p} \psi^{(p)}(x)\,\psi^{(p)*}(\xp).
    \label{Basis_interact_pot_link}
\end{equation}
To construct basis elements,
we periodize ${ U(x , \xp) }$
on a period ${2L}$,
so that it becomes $\Uper(x , \xp) \!=\! U (x , \xp) $
for ${ |x \!-\! \xp| \leq L }$,
and $\Uper (x \!+\! 2 kL , \xp) \!=\! \Uper (x , \xp)$
for ${ k \!\in\! \mathbb{Z} }$.
Dropping the constant term,
the periodized potential, $\Uper$,
is decomposed in Fourier series via
\begin{align}
    \label{periodized_pot_decomposition}
    \Uper {} & (x,\xp)  = 
    -\frac{4GL}{\pi^{2}} \sum\limits_{\substack{p \textrm{ odd} \\ p>0}} \frac{1}{p^2} 
    \\
    \times {} &
    \big[ \cos \big( p \tfrac{\pi}{L} x \big) \cos \big( p \tfrac{\pi}{L} \xp \big) 
    +  \sin \big( p \tfrac{\pi}{L} x \big) \sin \big( p \tfrac{\pi}{L} \xp \big) \big].
    \nonumber
\end{align}
Following Eq.~\eqref{Basis_interact_pot_link},
the natural basis elements are then
\begin{equation}
    \psi^{(p)}_{\mathrm{even}}(x) = \frac{2\sqrt{GL}}{p \, \pi} 
    \cos \big[ p \tfrac{\pi}{L} x \big] ,
    \label{psi_cossinbasis}
\end{equation}
with ${ p \!>\! 0 }$ odd,
and their odd counterpart $\psi^{(p)}_{\mathrm{odd}}$ 
via ${ \cos \!\to\! \sin }$.
Following Eq.~\eqref{derivative_Poisson_eq},
the associated densities are
\begin{equation}
    \rho^{(p)}_{\mathrm{even}}(x) = 
    \frac{-\pi^{2} \, p^{2}}{2GL^{2}} \psi^{(p)}_{\mathrm{even}}(x),
    \label{rho_cossinbasis}
\end{equation}
and equivalently for the odd ones.
It is straightforward to check that Eqs.~\eqref{psi_cossinbasis} and~\eqref{rho_cossinbasis}
comply with Eq.~\eqref{bi_ortho_basis}
for the periodized potential, $\Uper$,
when restricting the integration range to
${ -L \!\leq\! x \!\leq\! L }$.

In practice, the basis elements
are computed from coupled recurrence relations~\citep[see Eq.~{(5.4.6)} in][]{Press2007}.
In the main text,
we use a periodization length $L\!=\!10\Lambda$ (resp.\ $L\!=\!100\Lambda$) 
and 256 (resp.\ 1024) basis elements for thermal (resp.\ Plummer) computations. Indeed,
since the Plummer equilibrium density has wide tails (see Fig.~\ref{fig:densityprofiles}),
a large $L$ is required which, in turn, requires more basis elements
to reach a sufficient resolution.

\subsection{Fourier transform in angles}
\label{subapp:FT}

Once a suitable bi-orthogonal basis has been constructed, one has to compute 
the \FT\@ of the basis element, ${ \psi_{k}^{(p)} (J) }$,
involved in both the 
response matrix from Eq.~\eqref{def_M} and the dressed 
coupling coefficient
\begin{equation}
    \psid_{k\kp} (J , \Jp , \omega) = 
    - \sum_{p , q} \psi_{k}^{(p)} (J) \, \big[ \bI \!-\! \bM (\omega) \big]^{-1}_{pq} (\omega) \, 
    \psi^{(q) *}_{\kp} (\Jp),
    \label{def_psid}
\end{equation}
with $\bI$ the identity matrix,
and ${ \bM (\omega) }$ the system's response matrix (Eq.~\ref{def_M}).
Given the convention from Eq.~\eqref{map_angle_part},
the \FT\@ 
of the basis elements reads
\begin{equation}
    \psi_{k}^{(p)} (J) = 
    \frac{1}{\pi} \!\! \int_{0}^{\pi} \!\! \rd \theta \, \psi^{(p)} 
    \big( x [ \theta , J ] \big) \cos (k\theta) .
    \label{basis_FT}
\end{equation}
To compute this integral,
we naturally perform the 
same change of variables as 
in Appendix \ref{subapp:AA}.
One is left with two integrals that must be performed simultaneously
\begin{subequations}
    \label{FT_joint_integrals}
    \begin{align}
        \label{FT_psi_integral}
        \psi^{(p)}_{k} (J) & = 
        \frac{1}{\pi}  \int_{-1}^{1} \!\! \rd u \, 
        \dd[u]{\theta} \, \psi^{(p)} (x[u]) \, \cos (k \, \theta [u]) ,
        \\
        \label{FT_theta_integral}
        \theta [u] & = \!\! \int_{-1}^{u} \!\! \rd\up \, \dd[\up]{\theta},
    \end{align}
\end{subequations}
where ${\rd\theta / \rd u \!=\! \Omega(\rapo) I(u,\rapo) / \sqrt{2}}$ 
with ${I(u,\rapo)}$ defined in Eq.~\eqref{integrand_anomaly}.
Although the integrals from Eqs.~\eqref{FT_joint_integrals} seem nested, 
they can be evaluated via
the single integral of a $2$-vector~\citep{Rozier+2019}. 
In practice, we use
a \RK\@ scheme with $10^{3}$ steps 
for ${u\in [-1,1]}$.

\subsection{Bare coupling coefficients}
\label{subapp:Multipole}

In the Landau case, collective effects can be neglected.
As such, in Eq.~\eqref{def_psid}, one makes
the replacement ${ [\bI - \bM (\omega)]^{-1} \!\to\! \bI }$,
and the dressed coupling coefficients, ${ \psi_{k\kp}^{\rd} (J , \Jp , \omega) }$
become the bare ones, ${ \psi_{k\kp} (J , \Jp) }$.
Importantly, these coefficients
can be computed without any basis expansion, as they are
the Fourier transform of the pairwise interaction w.r.t.\ the angle $\theta$~\citep{Chavanis2013}.
Using the effective anomaly ${u}$ from Appendix \ref{subapp:AA},
the frequency-independent bare coupling coefficients
become
\begin{equation}
    \psi_{k\kp}(J,\Jp) = \frac{1}{\pi^{2}}
    \!\! \int_{-1}^1 \!\! \rd u\, \rd\up \; g(x)\, \gp(\xp)\; U(x,\xp),
    \label{bare_psi_multipole}
\end{equation}
with ${g(x)\!=\! \cos(k\theta)\,\rd\theta/\rd u}$ (and similarly for 
$\gp$).
Symmetry imposes ${ \psi_{k\kp} (J,\Jp) \!=\! 0 }$
for any $k,\kp$ of different parity.
The same result also holds for the dressed susceptibility coefficients,
${ \psi_{k\kp}^{\rd} (J , \Jp , \omega) }$, from Eq.~\eqref{def_psid}.

To compute Eq.~\eqref{bare_psi_multipole},
each anomaly, $u,\up$,
is sampled with $K$ nodes
at the location ${ u_{i} \!=\! -1 \!+\! 2 (i \!-\!\half)/K }$
with ${ 1 \!\leq\! i \!\leq\! K }$.
Equation~\eqref{bare_psi_multipole} becomes 
\begin{equation}
    \psi_{k\kp} (J,\Jp) = \frac{4G}{\pi^{2}K^{2}} \sum_{i,j=1}^{K} g_i \, \gp_j \, |x_{i} - \xp_{j}| ,
    \label{bare_psi_samp}
\end{equation}
where the ${g_i\!=\!g(x_i) \!=\! g (x (u_{i}))}$
(and ${\gp_j}$) are pre-computed
in a single pass using a direct integration of ${\rd\theta / \rd u}$,
following Eq.~\eqref{FT_joint_integrals},
requiring $\mO (K)$ operations.

The quasi-separable form of the pairwise interaction potential allows us to
 rewrite Eq.~\eqref{bare_psi_samp}
as
\begin{equation}
    \psi_{k\kp} (J , \Jp) = 
    \frac{4 G}{\pi^{2}K^{2}} 
    \sum_{j = 1}^{K} \gp_{j} \, \big( P_{j} + Q_{j} \big) ,
    \label{expand_sum_multi}
\end{equation}
with the cumulative sums
\begin{equation}
    P_{j} = \sum_{i = 1}^{w_{j}} g_{i} \, \big( \xp_{j} - x_{i} \big),
    \;
    Q_{j} = \sum_{\mathclap{i = w_{j} + 1}}^{K} g_{i} \, \big( x_{i} - \xp_{j} \big),
    \label{def_P_Q}
\end{equation}
and ${w_{j} \!=\! \Card \big\{ i \in \llbracket 1,K \rrbracket \, \big| \, x_{i} \leq \xp_{j} \big\}}$.
Importantly, $P_{j}$ and $Q_{j}$
can both be computed in a single pass,
requiring overall ${ \mO (K) }$ operations
to estimate ${ \psi_{k\kp} (J , \Jp) }$.
In practice, we used ${ K \!=\! 10^{3} }$ nodes,
and an \RK\@ scheme to compute ${g_i,\gp_j}$.
We note that for ${ k \!\gg\! 1 }$,
 ${ \psi_{kk} (J , J) \!\propto\! 1/k^{2} }$,
which explains the minor role played
by high order resonances, as in Fig.~\ref{fig:ContribRes}.

\subsection{Computing the response matrix}
\label{subapp:RepMat}

The response matrix from Eq.~\eqref{def_M} involves 
a sum over the resonances ${k}$,
and an integral over the action ${J}$ 
with a resonant denominator.
This asks for a careful treatment.

Benefiting from the rapid decay of the coupling coefficients,
we can safely truncate the sum over $k$ to ${ |k| \!\leq\! \kmax }$.
In practice, ${ \kmax \!=\! 10 }$ proves highly sufficient.

To deal with the resonant integral from Eq.~\eqref{def_M},
we follow the approach from~\cite{FouvryPrunet2022}:
\begin{enumerate}[wide, labelindent=0pt, label=(\roman*)]
\setlength\itemsep{-0.1em}
    \item The truncated action domain
    $[J_{0} , J_L]$ (with ${J_{0}\!=\! J(\rapo \!=\! 0)}$
    and ${ J_L \!=\! J (\rapo \!=\! L)}$)
    is remapped to ${[-1,1]}$ via
    ${y = \Sign (k) (\Omega (J) \!-\! \Sigma_{\Omega})/\Delta_{\Omega}}$ 
    with ${\Sigma_{\Omega} \!=\! \half(\Omega_0 \!+\! \Omega_L)}$,
    ${\Delta_{\Omega} \!=\! \half (\Omega_0 \!-\! \Omega_L)}$, 
    ${\Omega_0\!=\!\Omega(J_0)}$ and ${\Omega_L\!=\!\Omega(J_L)}$.
    Equation~\eqref{def_M} then becomes 
    \begin{equation}
        M_k^{pq}(\omega) = \!\! \int_{-1}^{1} \!\! \rd y \, \frac{G_k^{pq}(y)}{y - \varpi_k},
        \label{Mkpq}
    \end{equation}
    with
    \begin{subequations}
        \label{Gkpq_varpi}
        \begin{align}
            \label{Gkpq}
            G_k^{pq}(y) 
            &= 2\pi\, \Sign (k) \, \dd[\Omega]{J}
            \frac{\partial F}{\partial J}\psi_{k}^{(p) *} (J) \, \psi_{k}^{(q)} (J),
            \\
            \label{varpi}
            \varpi_k &= \frac{\omega}{|k|\Delta_{\Omega}} - \Sign (k) \frac{\Sigma_{\Omega}}{\Delta_{\Omega}},
        \end{align}
    \end{subequations} 
    where $J$ depends implicitly on $y$.
    \item The numerator ${G_k^{pq} (y)}$ in Eq.~\eqref{Gkpq}
    is projected onto Legendre polynomials
    via ${ G_{k}^{pq} (y) \!=\! \sum_{\ell=0}^{\ellmax} a_{k\ell}^{pq} \, P_{\ell} (y) }$,
    using a \GL\ quadrature
    truncated to $\ellmax$ (${ =\! 100 }$ in practice).
    Equation~\eqref{Mkpq} then becomes $M_k^{pq}(\omega) = \sum_{\ell=0}^{\ellmax} a_{k\ell}^{pq} \, D_{k\ell}(\omega)$ with
    \begin{equation}
        \label{D_ell}
        D_{k\ell}(\omega) =\!\! \int_{-1}^{1} \!\! \rd y \frac{P_{\ell}(y)}{y-\varpi_k}.
    \end{equation}
    \item We apply Landau's prescription~\citep[see, e.g.\@, Sec.\@ 5.2.4 in][]{BinneyTremaine2008} to compute 
    ${D_{k0}}$ and ${D_{k1}}$,
    while $D_{k\ell}$ for ${ \ell \geq 2 }$ are computed via
    direct recurrences~\citep[see Appendix D in][]{FouvryPrunet2022}. 
\end{enumerate}

\subsection{Quasi-stationary states}
\label{subapp:QSS}

The equilibrium \DFs\
presented in Sec.~\ref{subsec:GTE}
are obtained by
Eddington inversion~\citep[see, e.g.\@, Sec.\@ 4.3.1 in][]{BinneyTremaine2008}.
For a symmetric density profile,
the density ${ \rho (r \!=\! |x|) \!=\! 2 \!\int\!_{0}^{+\infty} \! \rd v \, F (E) }$,
can be expressed as
\begin{equation}
    \label{eq:rho_psi}
    \rho(\psi) = \sqrt{2} \!\! \int_\psi^{+\infty} \!\!\!\! \rd E \frac{F(E)}{\sqrt{E-\psi}} ,
\end{equation}
with $\psi = \psi (r)$.
Following Eq.~{(B.72)} of~\cite{BinneyTremaine2008},
this Abel integral equation is inverted as
\begin{equation}
    \label{eq:Eddington_inversion}
    F(E) = \frac{\sqrt{2}}{\pi} \!\! \int_E^{+\infty} \!\!\!\! \rd \psi \sqrt{\psi-E} \, \dds[\psi]{\rho}.
\end{equation}

Finally, using the relation ${ \psi (x) \!=\! \int\! \rd \xp \rho(\xp) U(x,\xp) }$,
one readily finds 
the potential of the thermal equilibrium
\begin{equation}
    \label{thermal_pot}
    \psi (x)= G \Mtot \Lambda \log 
    \left[2 \cosh \left( x / \Lambda \right)\right],
\end{equation}
as well for the Plummer quasi-stationnary equilibrium
\begin{equation}
    \label{Plummer_pot}
    \psi (x) = G \Mtot \alpha \, \sqrt{1+ (x / \alpha)^{2}}.
\end{equation}

The DF from Eq.~\eqref{eq:Thermal_DF} is the usual Boltzmann 
distribution $F(E)\propto e^{-mE/ \kB T}$ of statistical mechanics with
the specific energy $E$, and a thermodynamical temperature ${\kB T\!=\!mE_0/2}$. 
Using the virial theorem, one can relate the total energy $E_{\rm tot}$ to the 
temperature $T$, characteristic velocity $\sigma$ and characteristic length $\Lambda$ 
by  ${\Etot\!=\!\tfrac{3}{2}N\kB T\!=\!\tfrac{3}{4}\Mtot\sigma^2\!=\!\tfrac{3}{4}G\Mtot^2\Lambda}$. 

\begin{figure}[htbp!]
    \begin{center}
    \includegraphics[width=0.45\textwidth]{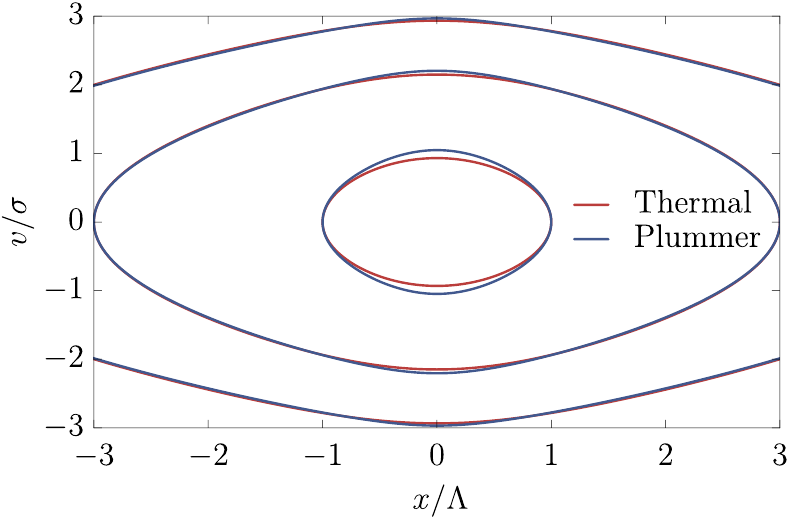}
    \caption{Typical mean-field (closed) orbits in phase-space,
    for apocenters ${\rapo/\Lambda\!=\!(1,3,5)}$.
    Because the Plummer equilibrium is more peaked than the thermal one, 
    its orbits reach a larger maximal velocity in the system's center.
    \label{fig:orbits}}
    \end{center}
\end{figure}

Figure~\ref{fig:orbits} illustrates typical mean-fields orbit
in the thermal and Plummer equilibria.
Both display similar phase space diagrams,
although Plummer's orbits reach larger central velocity 
owing to their denser core (Fig.~\ref{fig:densityprofiles}). 

\section{$N$-body integration}
\label{app:NBodyIntegration}

\subsection{Method}
\label{subapp:Method}

The system's total Hamiltonian is 
\begin{equation}
    \Htot = \sum_{i = 1}^{N} \half m_i v_i^2 + \sum_{i<j} m_i m_j \, U(x_i,x_j) ,
    \label{total_Hamiltonian}
\end{equation}
so that 
the equations of motion for particle ${i}$ read
\begin{equation}
        \dot{x}_{i} {} = v_{i} ;
        \quad
        \dot{v}_{i} {} =  G \left( M^{\rr}_{i} - M^{\rl}_{i} \right) ,
\label{EqsMotion}
\end{equation}
with $ {M^{\rr}_{i}}$ (resp.\ ${M^{\rl}_{i}}$) the total mass
on the right (resp.\ on the left) of particle ${i}$.
Importantly,
by sorting the set ${ \{ x_{i} \}}$,
one can compute these cumulative masses
in a single pass.
Determining the (exact) instantaneous forces on all particles
requires therefore ${ \mO (N \ln (N)) }$ operations.

The present \oneD\@ system can be integrated exactly
using a collision-driven scheme~\citep{Noullez2003}.
However, this approach requires ${ \mO(N^{2} \ln(N)) }$ operations
per dynamical time, making long-time integrations
of large-$N$ systems too challenging.
As such, we rather settle on using
an approximate time integrator
(with exact forces).
Because Eq.~\eqref{total_Hamiltonian} is separable,
one can use standard splitting methods~\citep[see, e.g.\@,][]{Hairer+2006}
to devise integration schemes.
The main source of error
comes from the abrupt force changes every time particles cross,
making it wiser to limit oneself to low-order schemes.
We use the standard leapfrog scheme~\citep[see, e.g.\@, Sec.\@ 3.4.1 in][]{BinneyTremaine2008}
which requires a single (costly) force evaluation
per timestep, ${ \delta t }$,
and an overall $\mO(N \ln (N) \, \tdyn / \delta t)$
operations per dynamical time.

\begin{figure}[htbp!]
    \begin{center}
    \includegraphics[width=0.48\textwidth]{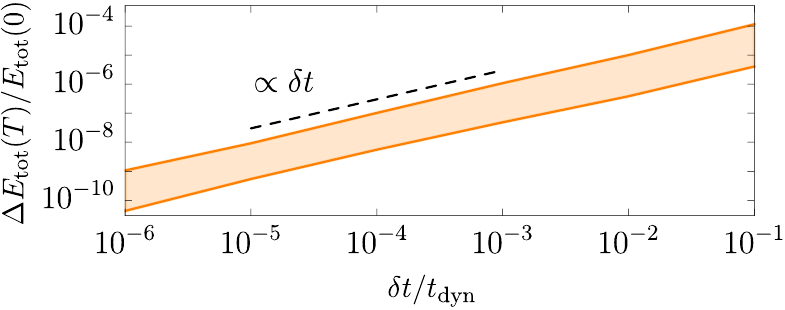}
    \includegraphics[width=0.48\textwidth]{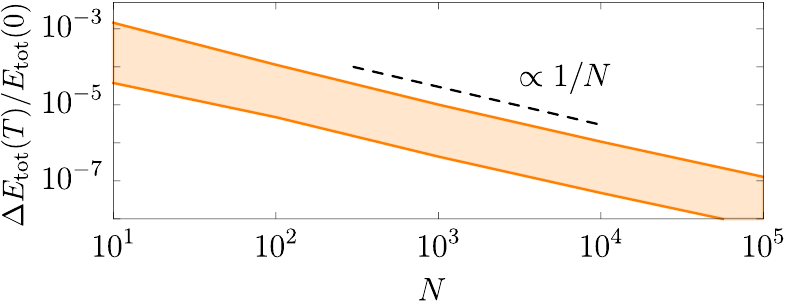}
    \includegraphics[width=0.48\textwidth]{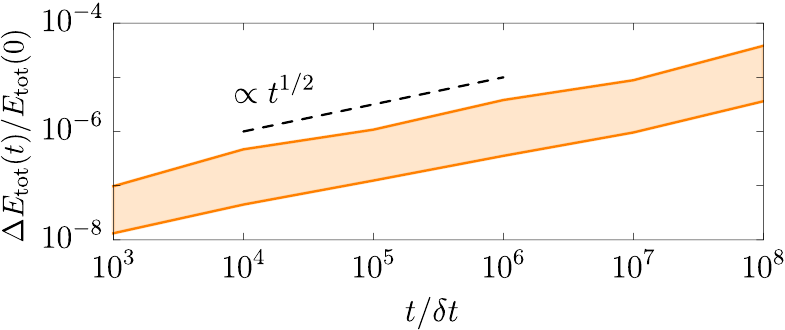}
    \caption{Relative error in the system's total energy, $\Etot$,
    as a function of
    (i) the timestep ${\delta t}$ (with ${N\!=\!10^{4}}$, ${T/\tdyn \!=\! 100}$),
    (ii) the number of particles $N$ (with ${\delta t / \tdyn \!=\!10^{-3}}$, ${T/\tdyn\!=\!100}$), 
    (iii) the total 
    number of integration steps ${t/\delta t}$ (with ${N\!=\!10^{4}}$, ${\delta t / \tdyn\!=\!10^{-3}}$).
    \label{fig:Nbody_convergence}}
    \end{center}
\end{figure}

In Fig.~\ref{fig:Nbody_convergence}, we check
the sanity of our algorithm,
by illustrating the conservation
of the total energy, $\Etot$, as one varies the timestep ${ \delta t }$,
the number of particles, $N$,
and the overall number of integration time steps, ${t/\delta t}$.
Because the pairwise interaction, ${ U(x,\xp) }$,
does not have a continuous derivative,
the leapfrog scheme is only first-order accurate,
i.e.\ its error scales like ${ \mO (\delta t) }$ after a fixed finite-time (top panel).
As one increases $N$, these discontinuities weaken,
so that the error at finite time scales like ${ \mO (1/N) }$
(center panel).
Finally, for the present explicit scheme,
we empirically find that the error in $\Etot$
grows like $\sqrt{t}$ as a function of time (bottom panel).

To prevent the $N$-body realizations
from drifting away, we systematically perform
the operation ${ v_{i} \!\leftarrow\! v_{i} \!-\! \sum_{i=1}^{N} \!m_{i} v_{i}/\Mtot }$
at the simulation's onset,
hence setting the system's total momentum to zero.
Such a recentring slightly blurs the effective \DF\@
in velocity space (and therefore in energy)
by an amount proportional to ${ 1/\sqrt{N} }$.
To mitigate this effect, we always chose values of $N$
large enough, e.g.\@, ${ N \!=\! 10^{5} }$ as in Fig.~\ref{fig:DEE}.

In the Landau simulations, 
we introduce two types of particles:
(i) massive background particles
that follow the smooth mean potential, and 
(ii) massless test particles driven
by the instantaneous (noisy) potential
generated by the background particles.
The orbital diffusion undergone by these test particles 
corresponds to the (undressed) Landau diffusion.

\subsection{Diffusion measurements}
\label{subapp:diff_measure}

To estimate diffusion coefficients in $N$-body simulations,
we follow Eq.~\eqref{diffusion_coefficient_interpretation}.
First, for the sake of convenience,
we measure diffusion in  energy,
${ E \!=\! v^{2}/2 \!+\! \psi (x) }$,
computed with ${ \psi(x) }$ the system's
initial unperturbed potential.
For a given realization, particles are initially binned
in 25 bins of width
${\dEbin\!=\!0.1 \, E_{0}}$
starting at the minimal energy ${\psi(0)}$.
For every bin and every time dump, we compute
${ \langle \Delta E^{2} (t) \rangle \!=\! \langle (E (t) \!-\! E (t\!=\!0))^{2} \rangle }$,
averaged over all the particles initially in the bin
and all the available realizations.
In practice, the associated time series, ${ t \!\to\! \langle \Delta E^{2} (t) \rangle }$
is truncated at a time $\Tmax$ chosen so that
${ \langle \Delta E^{2} (\Tmax) \rangle \!\leq\! \dEbin^{2} }$.
This ensures that particles have not diffused so much
as to explore too different energies.

Because the system's fluctuations are correlated,
the series of ${ \langle \Delta E^{2} \rangle }$
are not always linear function of time,
but exhibit initially a quadratic dependence w.r.t.\ time.
This occurs during the ballistic time, $\Tbal$,
which, fortunately is independent of $N$
(see Fig.~\ref{fig:TC_thermal}).
It is important not to perform
any measurement within this early phase.
A final caveat stems from the fact
that at large time, the \BL\@ time series
become sub-linear, a phenomenon already noted
in the \HMF\@ model~\citep[see Fig.~{8} in][]{Benetti+2017}.
This is accounted for by appropriately reducing
the series' maximal time, $\Tmax$,
so as not to enter this regime.

\begin{figure}[htbp!]
    \begin{center}
    \includegraphics[width=0.48\textwidth]{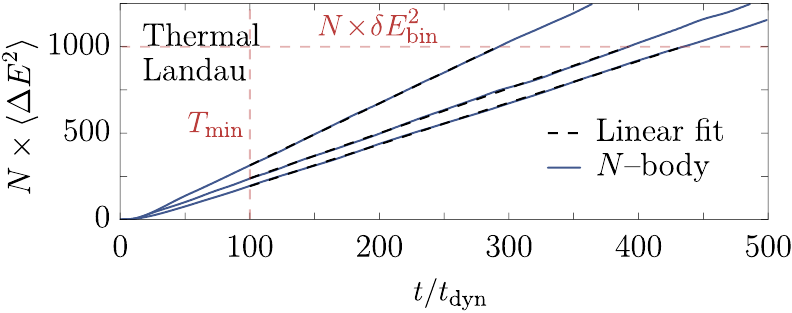}
    \includegraphics[width=0.48\textwidth]{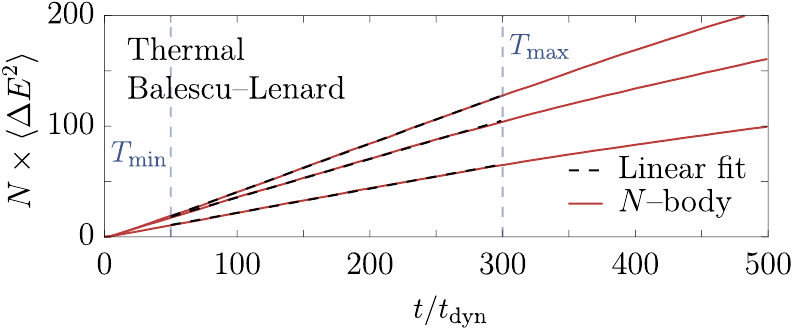}
    \caption{Typical time series of energy dispersion
    averaged over a given energy bin and ${1\,280}$ realizations
    for Landau (top panel) and \BL\@ measurements (bottom),
    together with the associated linear fit. 
    Here, ${\langle \Delta E^2 \rangle}$
    first evolves quadratically in time
    (ballistic regime)
    and then linearly (diffusive regime).
    For the \BL\@ experiments, the time series
    ultimately becomes sublinear,
    as already noted in the \HMF\@ model~\citep{Benetti+2017}.
    \label{fig:diff_fits}}
    \end{center}
\end{figure}

Once the domain ${ \Tbal \!\leq\! t \!\leq\! \Tmax }$
determined, we rely on Eq.~\eqref{diffusion_coefficient_interpretation}
and estimate the diffusion coefficient
with a linear fit (least squares) on that timespan.
This is illustrated in Fig.~\ref{fig:diff_fits}
for both Landau and \BL\@ measurements.

For the \BL\@ measurements in Fig.~\ref{fig:DEE},
we ran $10$ independent groups of ${1\,280}$ realizations
with ${N \!=\! 10^5}$ particles,
with ${ \delta t \!=\! 10^{-3} \, \tdyn }$
up to ${T \!=\! 500 \,\tdyn }$,
reaching a typical relative error in $\Etot$
of order $10^{-6}$,
and dumping ${\Delta E^2}$ values every $\tdyn$.
As illustrated in Fig.~\ref{fig:diff_fits},
we performed the linear fit within the domain
${[\Tbal,\Tmax]\!=\![50,300]\!\times\!\tdyn}$.
In Fig.~\ref{fig:DEE}, we report
the mean value and standard deviation
of the $10$ independent batches of realizations.

For the Landau experiments,
we use the exact same parameters,
except that the ${ N \!=\! 10^{5} }$ massive background particles
follow the smooth mean potential,
and we injected ${ 2 \!\times\! 10^{4} }$ massless test particles
sampled initially according to ${ F (E) }$.
Because Landau simulations exhibit longer correlation times
(see Fig.~\ref{fig:TC_thermal}),
we use ${\Tbal\!=\!100 \, \tdyn}$ and adjusted ${\Tmax}$
for every bin so that ${ \langle \Delta E^{2} (\Tmax) \rangle \!\leq\! \dEbin^{2} }$,
as illustrated in Fig.~\ref{fig:diff_fits}.

\subsection{Flux measurements}
\label{subapp:flux_measure}

To estimate the diffusion flux from Eq.~\eqref{BL_eq_def_flux},
we rely on the easy to measure cumulative \DoS
\begin{equation}
    \label{eq:cum_DoS_E}
    G(E) = \!\!\int_{-\infty}^E\!\! \rd E' \, P(E'),
\end{equation}
with ${P(E)\!=\!2\pi F(E) / \Omega (E)}$ the \DoS\@ in energy,
normalized so that ${ \!\int\! \rd E \, P (E) \!=\! 1 }$.
Following Eq.~\eqref{BL_eq_flux}, we naturally have
${ \rd G / \rd t \!=\! - 2 \pi \Flux(E) }$.
Consequently, to measure the flux,
we simply count the number of particles
with an (unperturbed) energy smaller than a given energy
threshold $E$, and keep track of this quantity as a function of time.
Once averaged over realizations,
the flux, ${ \Flux(E) }$, is directly estimated via linear fits.
These measurements are more challenging
than that of the diffusion coefficients
because the Plummer flux is particularly small (Fig.~\ref{fig:Flux}).

In pratice, we ran $10$ independent groups of ${1\,280}$ realizations
with ${ N \!=\! 10^4}$ particles,
with ${ \delta t \!=\! 10^{-3} \, \tdyn }$
up to ${ T \!=\! 10^{4} \, \tdyn}$,
reaching a typical relative error on $\Etot$
of order $10^{-5}$,
and dumping values of interest every ${ 10 \, \tdyn }$.
In Fig.~\ref{fig:Flux}, we report the mean value
and standard deviations over these $10$ independent batches.

\subsection{Correlation measurements}
\label{subapp:TC_measure}

As emphasized in~\cite{Binney+1988},
orbital diffusion is generically sourced by the time
correlation of the potential fluctuations,
which here stem from Poisson shot noise.
The instantaneous density
${\rho_{\rd}(x,t) \!=\! \sum_i m \deltaD (x \!-\! x_i(t))}$
can easily be projected onto the biorthogonal basis (Appendix \ref{subapp:BOB})
to write ${\rho_{\rd}(x,t) \!=\! \sum_p A_p(t) \rho^{(p)}(x)}$
with 
\begin{equation}
    A_p(t) = - \sum_i m \,\psi^{(p)}(x_i(t)) .
    \label{fluctuation_basis_coefficients}
\end{equation}
We use these coefficients to probe
the time evolution of the system's finite-$N$ fluctuations
In Fig.~\ref{fig:TC_thermal},
we illustrate the correlation
\begin{equation}
    C(t) = \!\!\int_{0}^{T-t} \!\!\!\! \frac{\rd\tau}{T-t} \langle  A_p(\tau) A_p(\tau+t)  \rangle,
    \label{eq:Time_correlation}
\end{equation}
for the odd basis element ${\psi_{\mathrm{odd}}^{(3)}}$,
following Eq.~\eqref{psi_cossinbasis}.
In practice, we ran ${1\,280}$ realizations
of the thermal equilibrium
with ${N \!=\! 10^5 }$ particles,
with ${ \delta t \!=\! 10^{-3} \, \tdyn }$
up to ${ T \!=\! 10^{3} \, \tdyn }$,
reaching a typical relative error on $\Etot$
of order $10^{-6}$,
and dumping values of ${A_p}$ every ${ 0.05 \, \tdyn }$.
For the \BL\@ experiment,
we also let the system ``warm up'' during $200$ dynamical times
before any measurement,
so as to let the initial Poisson shot noise thermalize
and get dressed by collective effects~\citep[see, e.g.\@, Appendix F in][]{Fouvry2018}.

\end{document}